%% file: _06Akalu_revision.tex
\documentclass[a4paper]{jpconf}
\usepackage{lmodern}        
\usepackage[T1]{fontenc}
\usepackage[utf8]{inputenc} 
\usepackage{graphicx}  
\usepackage{appendix}

\usepackage{newtxtext,newtxmath}
\usepackage{pgfplots}
\pgfplotsset{compat=1.18}
\usepackage{placeins} 

\usepackage[T1]{fontenc}
\usepackage{graphicx}
\usepackage{float} 
\usepackage{lipsum}
\DeclareRobustCommand{\VAN}[3]{#2}
\let\VANthebibliography\thebibliography
\def\thebibliography{\DeclareRobustCommand{\VAN}[3]{##3}\VANthebibliography}

\usepackage{graphicx}	
\usepackage{amsmath}	
\usepackage{graphicx}
\usepackage{float} 

\newcommand\Tstrut{\rule{0pt}{2.6ex}}        
\newcommand\Bstrut{\rule[-1.4ex]{0pt}{0pt}}  
\usepackage[greek, english]{babel}
\usepackage{alphabeta}
\usepackage{textgreek}

\usepackage[pdftex,bookmarksopen,bookmarksnumbered,linktoc=all,hidelinks]{hyperref}


\begin{document}
\vspace{-10.cm}
\title{Constraining the modified symmetric teleparallel gravity using cosmological data }

\author{Shambel Sahlu$^{1,2}$, and  Amare Abebe$^{1,3}$}

\address{$^1$ Centre for Space Research, North-West University, Potchefstroom 2520, South Africa}
\address{$^2$ Department of Physics, Wolkite University, Wolkite, Ethiopia}
\address{$^3$ National Institute for Theoretical and Computational Sciences (NITheCS), South Africa}

\ead{\url{shambel.sahlu@nithecs.ac.za}}
\ead{\url{Amare.Abebe@nithecs.ac.za }}
\begin{abstract}
This paper examines the late-time accelerating Universe and the formation of large-scale structures within the modified symmetric teleparallel gravity framework, specifically using the \( f(Q) \)-gravity model, in light of recent cosmological data. After reviewing the background history of the Universe, and the linear cosmological perturbations, we consider the toy model \( F(Q) = \alpha\sqrt{Q}+\beta   \) ( where \( Q\) represents nonmetricity, \(\alpha\)  and \(\beta\) are model parameters) for further analysis. 
To evaluate the cosmological viability of this model, we utilize 57 Observational Hubble Data (\texttt{OHD}) points, 1048 supernovae distance modulus measurements (\texttt{SNIa}), their combined analysis (\texttt{OHD+SNIa}), 14 growth rate data points (\texttt{f}-data), and 30 redshift-space distortions (\texttt{f}$\sigma_8$) datasets. Through a detailed statistical analysis, the comparison between our model and $\Lambda$CDM has been conducted after we compute the best-fit values through the Markov Chain Monte Carlo (MCMC) simulations. Based on the results, we obtain the Hubble parameter, $H_0 = 69.20^{+4.40}_{\text{--}2.10}$ and the amplitude of the matter power spectrum normalization $\sigma_8 = 0.827^{+0.03}_{\text{--}0.01}$. These values suggest that our model holds significant promise in addressing the cosmological tensions.
\end{abstract}

\section{Introduction}
{The Universe has experienced a late-time accelerated phase of expansion, and the reason for this has mostly been attributed to the influence of a yet unknown - exotic - form of energy called dark energy (DE). Late-time acceleration has been demonstrated by a variety of observation measurements, including Type Ia Supernovae \cite{riess1998observational,Perlmutter_1999}, large-scale structure (LSS) \cite{koivisto2006dark,daniel2008large}, Baryon Acoustic Oscillations (BAO) \cite{eisenstein2005detection,2011MNRAS.417.3101P}, and microwave background (CMB) anisotropies \cite{caldwell2004cosmic,huang2006holographic}, but so far no experiment or observation has made a direct detection of DE. As such, there have been several alternative proposals to explaining the late-time acceleration, including modified gravity theories (MGTs). Such a need for MGTs has, in the last decade or so, been more prominent due to the discovery of discrepancies - referred to as `` cosmological tensions" - in local and early-universe measurements of certain cosmological parameters, like the Hubble constant $H_0$ or the large-scale structure parameter $\sigma_8$. Among the most widely explored MGTs include the matter-geometry coupling expressed as $f(R)$ gravity ($R$ being the curvature scalar) \cite{harko2011f}, $f(T)$ gravity (${T}$ being the torsion scalar) \cite{setare2013can, sahlu2020scalar}, $f(R, L_m)$  gravity, where $L_m$ is the matter Lagrangian density \cite{sahlua2024cosmology},  $f(Q)$ gravity, where $Q$ is the scalar of non-metricity. }
\\
\\
The $f(Q)$-gravity model is a modification of general relativity where the Einstein-Hilbert action is replaced by a more general function of the non-metricity scalar $Q$ \cite{heisenberg2024review,mandal2020cosmography,atayde2021can}. Non-metricity refers to the metric tensor $g_{\alpha_\beta}$ not being covariantly conserved. Recently, \( f(Q) \)-gravity has been given significant attention in different aspects of cosmological problems, particularly for its potential to address some of the challenges faced by the standard \(\Lambda\)CDM model. In contrast to general relativity (GR), which is based on the curvature of spacetime, \( f(Q) \)-gravity is formulated based on the concept of non-metricity \cite{atayde2021can}, where the gravitational interaction is governed by the non-metricity scalar \(Q\), describing how distances and angles change under parallel transport without invoking curvature or torsion \cite{jimenez2018coincident}. By introducing a general function \( f(Q) \), this theory modifies the Einstein field equations, leading to a richer set of dynamics that can influence the evolution of the Universe in ways that differ from the standard model. The cosmological implications of \( f(Q) \)-gravity are far-reaching, including its impact on the Universe's expansion history, the growth of large-scale structures \cite{sokoliuk2023impact}, and the anisotropies observed in the cosmic microwave background (CMB) \cite{frusciante2021signatures,atayde2023f}. Furthermore, the theory offers new perspectives on the nature of dark energy and dark matter, with its extra degrees of freedom potentially serving as an effective fluid with properties distinct from those in \(\Lambda\)CDM. Moreover, the behaviour of gravitational waves within this framework provides another avenue for testing the theory against observations. However, while \( f(Q) \)-gravity holds promise, it also faces challenges, such as ensuring stability, fitting observational data, and remaining consistent with solar system tests of gravity. Ongoing and future research will be crucial in determining whether \( f(Q) \)-gravity can serve as a viable alternative to the standard cosmological model or if it will be constrained by empirical evidence. In this paper, the Universe's accelerating expansion and the growth of cosmic structures are highlighted by constraining cosmological parameters and conducting an in-depth analysis within the framework of $f(Q)$ gravity models using cosmological data. The statistical analysis has been performed after we computed the constraining parameters using the corresponding data for comparison between  \( f(Q) \) gravity and \(\Lambda\)CDM models by applying the Akaike Information Criterion (AIC) and the Bayesian/Schwarz Information Criterion (BIC), with the \(\Lambda\)CDM model serving as a baseline to evaluate the viability of the \( f(Q) \) models based on the work in \cite{szydlowski2015aic}. Our findings indicate that the  \(\Delta\)AIC values and the \(\Delta\)BIC values are indicating the \( f(Q) \) gravity model demonstrates significant observational support for the considered datasets. 
\\
The paper is organised as follows: Sect. \ref{theory} reviews the theoretical framework of $f(Q)$ gravity cosmology, presenting key mathematical expressions for background quantities and the evolution equation for density fluctuations. Sect. \ref{datameth} outlines the methods and data used in the analysis. Sect. \ref{resultdiscussion} presents and discusses the results, including a comparison of the $f(Q)$ gravity model with $\Lambda$CDM to test the viability of $f(Q)$ gravity in studying late-time cosmic expansion and large-scale structure formation. This section also includes diagrams of the Hubble parameter \(H(z)\), the distance modulus \(\mu(z)\), growth rate, and redshift-space distortion $f\sigma_8(z)$ that is a powerful tool in observational cosmology, to study the large-scale structure of the universe and probes of the structure growth. Finally, Sect. \ref{conclusions} offers the conclusions of the study.

\section{Theory}\label{theory}
In this section, the theoretical framework of $f(Q)$-gravity is highlighted, which is fundamental to studying the late-time cosmic acceleration and the formation of large-scale structures. We start by employing the formalism presented in \cite{jimenez2018coincident}, with the action given as:

\begin{equation}\label{eq1}
	S=\int\sqrt{-g} \left(\frac{1}{2}f(Q)+\mathcal{L}_m\right) {\rm d}^4x\;,
\end{equation}
where $f(Q)$ is an arbitrary function of the non-metricity $Q$, $g$ is the determinant of the metric \(g_{\mu\nu}\), and \(\mathcal{L}_m\) is the matter Lagrangian density. The energy-momentum tensor can be expressed as $T_{\mu\nu} = -\frac{2}{\sqrt{-g}}\frac{\delta(\sqrt{-g}\mathcal{L}_m)}{\delta g^{\mu\nu}}$. 
 By assuming a spatially flat Friedmann-Lema\^{i}tre-Robertson-Walker (FLRW) metric
\begin{equation}
    {\rm d}s^2 = -{\rm d}t^2 +a^2(t)\delta_{\alpha\beta}{\rm d}x^\alpha {\rm d}x^\beta\;,
\end{equation}
where $a(t)$ is the cosmological scale factor,\footnote{$\delta^{\alpha}{}_{\beta} =1$, when $\alpha = \beta$, zero otherwise.} the trace of the non-metric tensor becomes $Q = 6H^2$.  The corresponding modified Friedmann equation once the gravitational Lagrangian has been split into $f(Q) = Q +F(Q)$ becomes
\begin{eqnarray}\label{friedman}
	3H^2= \rho + \rho_r+\rho_{Q} \;,
\end{eqnarray}
where $\rho$, $\rho_r$ and $\rho_{Q}$ are the energy density of non-relativistic matter, relativistic matter (radiation), and dark energy from the contributions of $f(Q)$ gravity respectively. {{The energy density and the pressure terms coming from the non-metricity contribution of the $f(Q)$ gravity model read as
  \begin{eqnarray}\label{pressure}
	&&\rho_{Q}=\frac{F}{2}-QF',\label{55}~~\mbox{and}~~p_{Q} = -\rho_{Q}+2\dot{H} (2QF''+F')\;. \label{555}
\end{eqnarray}
{For further investigation, we consider a specific form of $f(Q$ model chosen in such a way that it mimics $\Lambda$CDM~\cite{frusciante2021signatures} and has as fewer free parameters as possible. One such model worth exploring is the $F(Q) = \alpha \sqrt{Q} +\beta$ toy model. } For the case of $\alpha =0$ and an appropriate choice of $\beta$, our model reduces to $\Lambda$CDM, and for the case of $\alpha = \beta = 0$, it reduces to pure GR.}}  To avoid any dimensional problem we assume $\beta = b_0 H_0^2$, where $b_0$ is a free parameters.  Then, from Eq. \eqref{friedman} and \eqref{555}, the Hubbel parameter $H(z)$   can be expressed as
\begin{eqnarray}\label{Hubbel}
   H(z) =  H_0\sqrt{\Omega_m (1 + z)^3 +b_0}\;.
\end{eqnarray}

The general form of distance modulus \(\mu(z)\) 
\begin{equation}\label{distancemodules1}
	\mu(z) =  25+5\times\log_{10}\left[3000\bar{h}^{-1}(1+z)\int^{z}_{0}\frac{dz}{h(z)}\right]\;,
\end{equation}
has been considered for the further investigation on the viability of the $f(Q)$ gravity models with SNIa data to study the acceleration of the expansion of the Universe,\footnote{This distance modulus is given in terms of {Mpc}.} where $\bar{h} = H_0/100$ and \(h(z) = H(z)/H_0\) is the normalised Hubble parameter for $f(Q)$ gravity model.
\\
For the cosmological background, we assume a homogeneous and isotropic expanding Universe, focusing on the spatial gradients of gauge-invariant variables such as
\[
    D^m_a=\frac{a}{\rho_m}\tilde{\nabla}_a\rho_m\;,
	Z_a=a\tilde{\nabla}_a\theta\;,~~~\mathcal{W}_a=a\tilde{\nabla}_aQ,~~\mbox{and}~~\mathcal{L}_a=a\tilde{\nabla}_a\dot{Q}, 
\]
which represent the energy density, the volume expansion of the fluid \(\theta \equiv 3H\), the characterising the fluctuations in nonmetricity density ($\mathcal{W}_a$) and momentum \(\mathcal{L}_a\) \cite{dunsby1992cosmological,sahlua2024structure}.
{
 The work in \cite{sahlua2024structure} thoroughly examines scalar and harmonic decomposition techniques within the framework of the \( f(Q) \) gravity model and provides the evolution equation of the density contrast as \footnote{Based on the work in  \cite{sahlua2024structure}, the scalar gradient vraibles read as  \begin{equation}
		\delta_m=a\tilde{\nabla}^aD^m_a,\hspace{0.1cm}	Z=a\tilde{\nabla}^aZ_a,\hspace{0.1cm}\mathcal{W}=a\tilde{\nabla}^a\mathcal{W}_a,\hspace{0.1cm}\mathcal{L}=a\tilde{\nabla}^a\mathcal{L}_a.
	\end{equation}}
  \begin{eqnarray}
		&&\ddot{\delta}_m^k= -\left[ \frac{2\theta}{3}+\frac{\dot{Q}F'}{2Q}+\dot{Q}F''-w\theta\right]\dot{\delta}_m^k+\Bigg[w F+\theta\dot{Q}F''w  -\frac{\theta\dot{Q}F'w}{2Q}+\frac{(1+3w)\rho_m}{2}(1-w)-\frac{k^2}{a^2} w\Bigg]\delta_m^k \nonumber\\&&+\Bigg[ \frac{1}{2}F'   +QF'' \theta\dot{Q}F''' +\frac{\theta\dot{Q}F''}{2Q}-\frac{\theta\dot{Q}F'}{2Q^2}\Bigg](1+w)\mathcal{W}^k  +\left[ \frac{\theta F'}{2Q} -\theta F''\right](1+w)\dot{\mathcal{W}}^k\,, \label{70011}\\&&
		\ddot{\mathcal W}^k=\frac{\dddot{Q}}{\dot{Q}}\mathcal W^k-\frac{2w \ddot{Q}}{1+w}\delta^k_m-\frac{w \dot{Q}}{1+w}\dot{\delta}_m^k\,.\label{700x}
	\end{eqnarray}
 As widely considered the quasi-static approximation in various modified gravity studies such as \cite{sahlu2020scalar, abebe2013large} we assumed that the in first and second-order time derivatives of non-metric density fluctuations are approximately zero (\(\dot{\mathcal{W}} = \ddot{\mathcal{W}} \approx 0\)). By taking into account the given model $f(Q) = \alpha\sqrt{Q}+\beta$, the simplified form of the redshift-dependent density contrast equation is obtained from Eqs. \eqref{70011} - \eqref{700x}  as
\begin{eqnarray*}
   && \frac{d^2\delta_m}{dz^2} =    \left(\frac{1}{1+z}- \frac{dE}{Edz}\left(1-\Omega_m-2b_0\right) \right)\frac{d \delta_m}{dz}+\frac{\Omega_{m}}{2E^2}(1+z)\delta_m\;.
\end{eqnarray*}
 The cosmic growth rate \( f(z) \), which quantifies the structure's growth, is a critical parameter in various observational studies, including redshift-space distortions \(\texttt{f}\sigma_8\). Derived from the matter density contrast \(\mathcal{\delta}_m\), the growth rate \( f(z) \) is expressed as \cite{springel2006large}:
\begin{eqnarray}
  && f \equiv \frac{{\rm d}\ln{{{\delta}}_m}}{{\rm d}\ln{a}} = -(1+z)\frac{\delta'_m(z)}{\delta_m(z)}
\;.\label{growth1}
\end{eqnarray}
 By admitting the definition of \eqref{growth1}, the evolution of the growth rate is governed by the following expression
\begin{eqnarray}\label{growthratelcdm}
    && (1+z)f' = f^2 -\left[(1+z)\frac{dE}{Edz}(1-\Omega_m-2b_0)-2\right] f -\frac{3  \Omega_m}{2E^2}(1+z)^3 \label{rr1}\;. 
\end{eqnarray}
 The redshift-space distortion \( f\sigma_8 \) is also obtained by combining the growth rate \( f(z) \) with the root mean square normalisation of the matter power spectrum, \( \sigma_8 \), measured within a sphere of radius \( 8h^{-1} \) Mpc:
\begin{eqnarray}\label{growth11}
  f\sigma_8(z)  = -(1+z)\sigma_{8,0}\frac{ {\rm d}\delta_m(z)}{{ \delta_m(z) \rm d}z}\;,
 \end{eqnarray}
where $\sigma_{8,0}$ is the present-day values of the root-mean-square normalisation of the matter power spectrum.}
\section{Data}\label{datameth}
In this study, we use the recent cosmic measurement datasets including: i)  Hubble parameter \(H(z)\) measurements (\texttt{OHD}), which consist of 57 data points \cite{yu2018hubble,dixit2023observational}  within the redshift range \(0.0708 < z \leq 2.36\), ii) Pantheon compilation of supernovae data (\texttt{SNIa}) \cite{scolnic2018complete}, which includes the 1048 distinct SNIa with redshifts ranging from \(z \in [0.001, 2.26]\) and iii) the combined dataset (\texttt{OHD+SNIa}). We also use the large-scale structure measurement namely: i) the growth rate (\texttt{f}) within the redshift range \(0.001 \leq z \leq 1.4\) and ii) redshift space distortion (\texttt{f}$\sigma_8$) covering the redshift range \(0.001 \leq z \leq 1.944\), to perform a thorough observational and statistical analysis \cite{avila2022inferring}. Additionally, we employed various software tools and Python packages, such as EMCEE \cite{hough2020viability} and GetDist (\cite{lewis2019getdist}), to constrain the model parameters \(\Omega_m, H_0,b_0\) using these cosmological datasets.
\section{Results and discussion}\label{resultdiscussion}
This section aims to present the detailed implications of the considered $f(Q)$ gravity model to broadly discuss the accelerating expansion of the late-time of the Universe and the formation of the growth structure.  After constraining the cosmological parameters \(\{\Omega_m, H_0\}\) and the exponent \(b_0\) using recent data via MCMC simulations, a detailed statistical analysis is conducted to test the viability of the $f(Q)$ gravity model against the $\Lambda$CDM. The results are discussed in detail in the following two subsections using the cosmic measurement and the large-scale structure measurements. 
\subsection{Using cosmic measurements}
\begin{figure}[h!]
\includegraphics[width=0.3\linewidth]{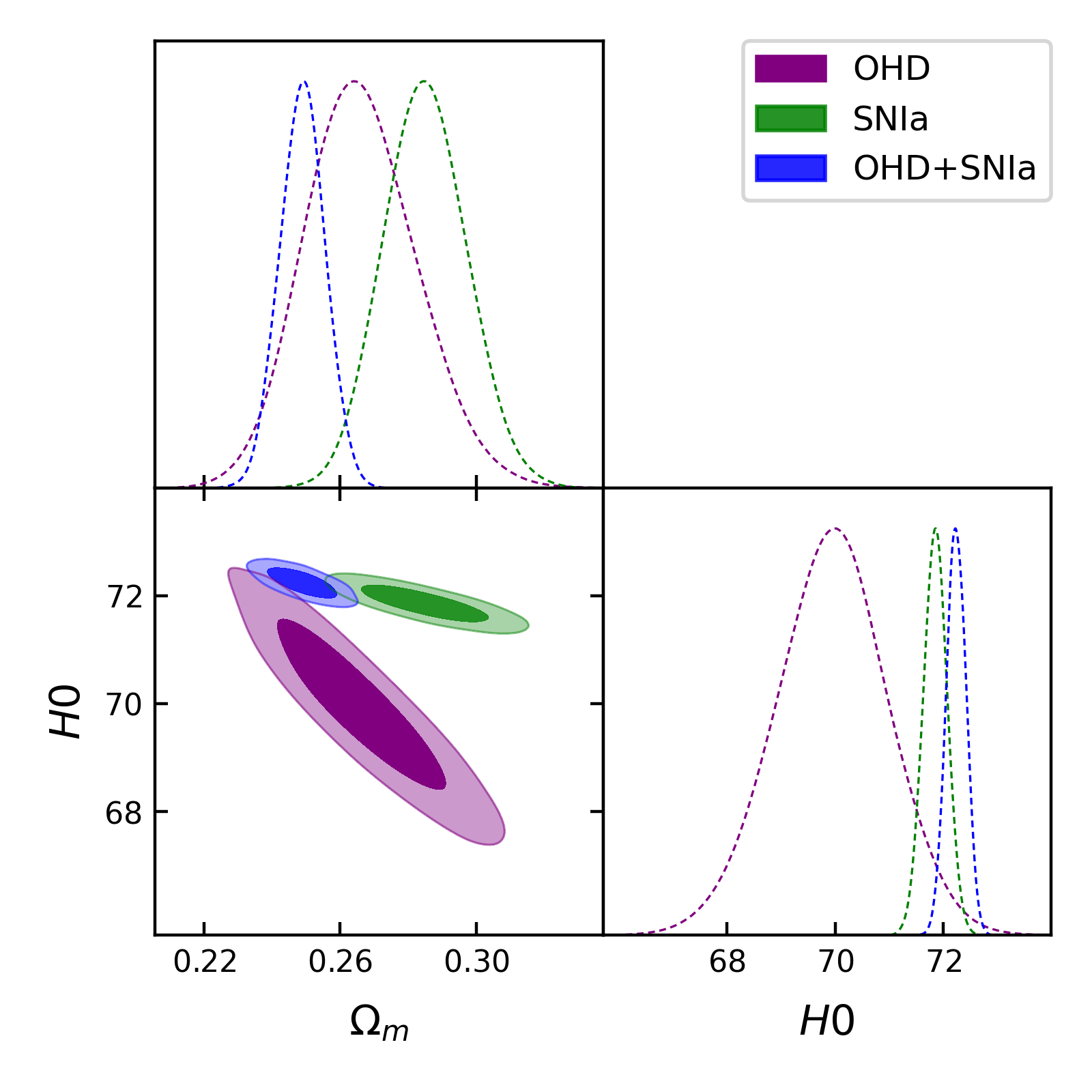}
~~~~~~~~~~~~~~~~~~~~~~~~~~~~~~~~~~~~~
    \includegraphics[width=0.3\linewidth]{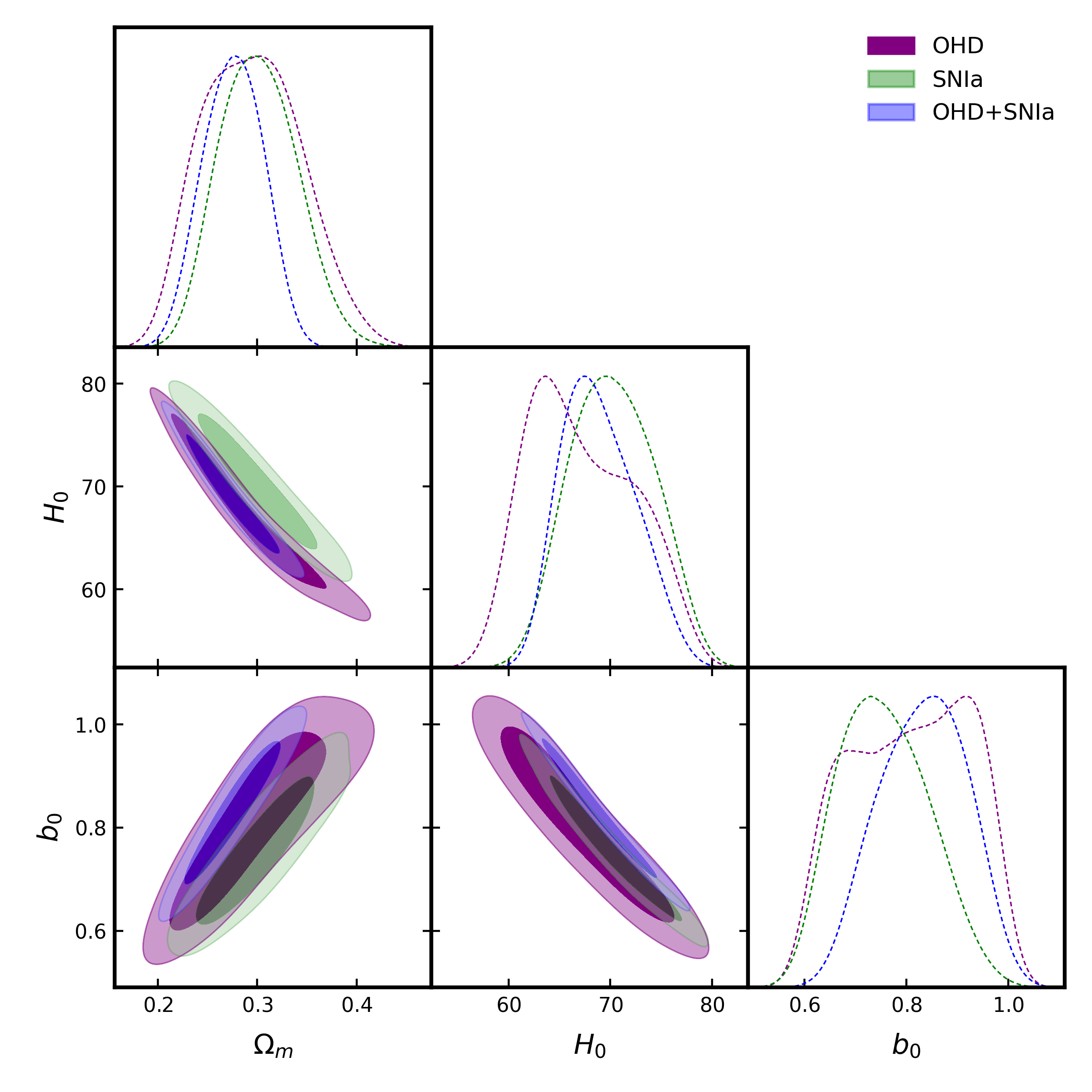}
    \caption{The best-fit values of the constraining parameters for the $\Lambda$CDM (left panel) and $f(Q)$ gravity model (right panel) are provided, using \texttt{OHD}, \texttt{SNIa}, and \texttt{OHD+SNIa} datasets from Eqs. \eqref{Hubbel} and \eqref{distancemodules1}. These values are shown at the $1\sigma$ and $2\sigma$ confidence levels.}
    \label{fig:enter-label}
\end{figure}

\begin{figure}[h!]
\centering
    \includegraphics[width=0.45\linewidth]{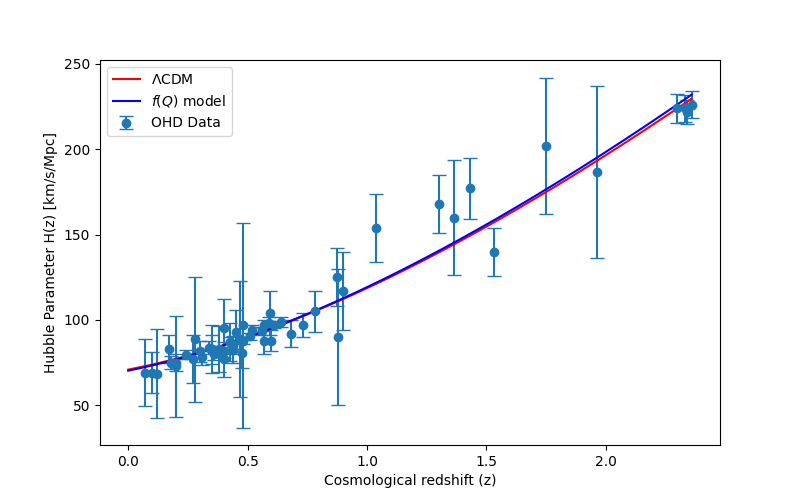}
    \includegraphics[width=0.45\linewidth]{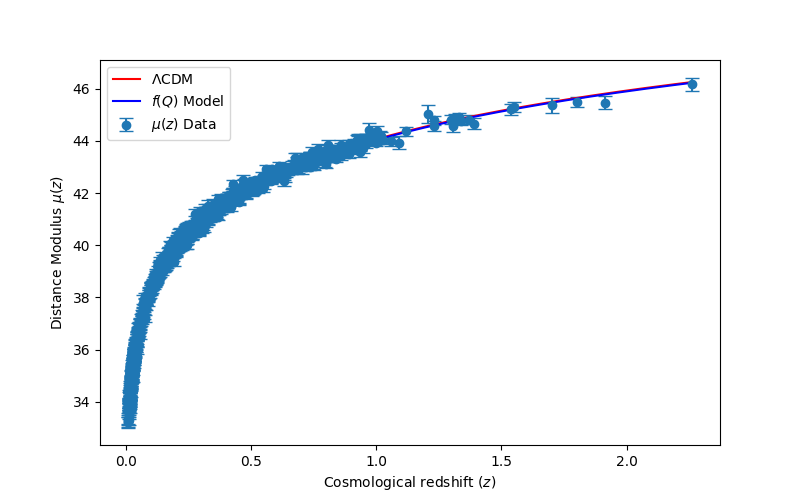}
    \caption{The diagram of the Hubble parameter \(H(z)\)  and the distance modulus \(\mu(z)\) as functions of redshift for the $\Lambda$CDM and $f(Q)$-gravity model using Eqs. \eqref{Hubbel} and  \eqref{distancemodules1}.}
    \label{fig:enter-label2}
\end{figure}

As broadly discussed in \cite{staicova2022constraining}, the constraining values of the cosmological parameters are key to understanding the evolution and composition of the Universe. After comparing the $f(Q)$ model with observational data, we constrain the $f(Q)$ gravity model to better understand the Universe's history and address key cosmic puzzles. As shown in Fig. \ref{fig:enter-label}, the parameter values constrained by MCMC simulations are presented with $1\sigma$ and $2\sigma$ confidence intervals, with further details provided in Table \ref{sphericcase1} for different datasets. 
From the contour plot in Fig. \ref{fig:enter-label}, we observe that the $\Lambda$CDM model (left panel) is inconsistent with the recent cosmic measurements at both the $1\sigma$ and $2\sigma$ levels. However, the $f(Q)$ gravity model (right panel) demonstrates better consistency with the data, highlighting a significant advantage of our model over $\Lambda$CDM. Using the best-fit values, we also generated the Hubble parameter $H(z)$ and distance modulus $\mu(z)$ diagrams as functions of cosmological redshift in Figs \ref{fig:enter-label2} in the \(\Lambda\)CDM and \( f(Q) \) models. The results show the Universe's expansion history, measure cosmic distances, and provide evidence for the Universe's accelerating expansion. We also evaluate the \( f(Q) \) gravity models against the \(\Lambda\)CDM model using the AIC and the BIC, using the \(\Lambda\)CDM model as a benchmark to determine the viability of the \( f(Q) \) models. From our results, the values of AIC = \( \{36.53, 1039.681, 1091.418\}\) and BIC = \(\{40.616, 1049.590,1101.433\}\) for $\Lambda$CDM using the \texttt{OHD}, \texttt{SNIa}, and \texttt{OHD}+\texttt{SNIa} datasets.  For $f(Q)$ gravity model AIC = \( \{38.243, 1042.132, 1095.000\}\) and BIC = \(\{43.540, 1053.80,1105.290\}\) for using the same datesets. The corresponding values of $\Delta$AIC reads \(\{1.713,2.442,3.582 \}\) and   $\Delta$BIC becomes \(\{2.923,4.21, 3.85\}\) which indicates the $f(Q)$ gravity model has substantial observational support. Considering this analysis, we will consider this $f(Q)$ gravity model for further investigation using the large-scale structure data in the next section. 
\begin{table*}
\caption{The best-fit values for both the \(\Lambda\)CDM and \(F(Q) = \alpha \sqrt{Q} +\beta \) gravity models were determined using the \texttt{OHD}, \texttt{SNIa}, \texttt{OHD}+\texttt{SNIa}, \texttt{f} and \texttt{f}$\sigma_8$ datasets. }  
\label{sphericcase1}
\begin{tabular*}{\textwidth}{@{\extracolsep{\fill}}lrrrrrrrl@{}}
\hline
\textbf{Data }& Model & $\Omega_m$& $H_0$ &$b_0$&$\sigma_8$ \\
\hline
  & $\Lambda CDM$ &&&&\\
    \hline
    \texttt{OHD}  & & $0.266^{+0.015}_{\text{--}0.018}$ & $70.105^{+1.10}_{\text{--}1.10}$ &- & - \Tstrut\\

   \texttt{SNIa} & & $0.284^{+0.013}_{\text{--}0.013}$& $71.860^{+1.30}_{\text{--}1.30}$ &-&  - \Tstrut\\
   \texttt{OHD+SNIa} & & $0.249^{+0.066}_{\text{--}+0.066}$ & $72.230^{+0.18}_{\text{--}0.18}$ &- &  - \Tstrut \Bstrut\\
   \texttt{f} & & $0.266^{+0.012}_{\text{--}0.012}$ &- & - &-\Tstrut \Bstrut\\
   \texttt{f}$\sigma_8$ & & $0.243^{+0.009}_{\text{--}0.009}$ &- & - & $0.741^{+0.028}_{\text{--}0.028}$\Tstrut \Bstrut\\
			\hline
    & $f(Q)$-gravity& && & \\
   \hline
   \texttt{OHD}  & & $0.280^{+0.056}_{\text{--}0.056}$ & $69.2^{+4.40}_{\text{--}2.10}$ & $0.771^{+0.210}_{\text{--}0.098}$ &-\Tstrut\\
  \texttt{SNIa} & & $0.306^{+0.057}_{\text{--}0.057}$& $70.3^{+4.10}_{\text{--}3.80}$ & $0.780^{+0.190}_{\text{--}0.110}$   &-\Tstrut\\
   \texttt{OHD+SNIa} & & $0.254^{+0.070}_{\text{--}0.070}$ & $70.9^{+3.10}_{\text{--}4.10}$ & $0.797^{+0.180}_{\text{--}0.088}$ &-\Tstrut \Bstrut\\
   \texttt{f} & & $0.313^{+0.059}_{\text{--}0.089}$ &- & $0.283^{+0.025}_{\text{--}0.014}$ &-\Tstrut \Bstrut\\
   \texttt{f}$\sigma_8$ & & $0.336^{+0.056}_{\text{--}0.079}$ &- & $0.263^{+0.013}_{\text{--}0.007}$ & $0.827^{+0.030}_{\text{--}0.007}$\Tstrut \Bstrut\\
   \hline
\end{tabular*}
\end{table*}

\subsection{Using large-scale structure data}
The growth rate \( f(z) \) is essential for understanding the formation and evolution of cosmic structures, while \( f\sigma_8 \) is a critical observable that integrates this growth rate with the clustering amplitude, offering a powerful method for probing the dynamics of dark energy and testing gravitational theories. In this study, we comprehensively analyse structure growth using large-scale structure data, comparing the predictions of the $\Lambda$CDM and $f(Q)$ gravity models. In most studies \cite{avila2022inferring}, an approximation method is often applied to simplify the expression for the growth rate, leading to a relation like \( f(z) = -(1+z)\frac{\delta_m'(z)}{\delta_m(z)} \approx \bar{\Omega}_m^\gamma \), where \( \gamma \) is known as the growth index, and \( \bar{\Omega}_m= \Omega_m(1+z)^3/h^2(z) \) represents the matter density parameter. This approximation provides a more manageable form for analytical calculations. However, in this work, we deviate from the common practice by employing the full expression for \( f(z) \)  without applying the simplification, allowing for a more accurate and detailed analysis of the growth rate. In this approach, the growth rate is highly dependent on the energy density contrast as we can see in Eq. \eqref{rr1}.  
\begin{figure}[h!]
\centering
    \includegraphics[width=0.6\linewidth]{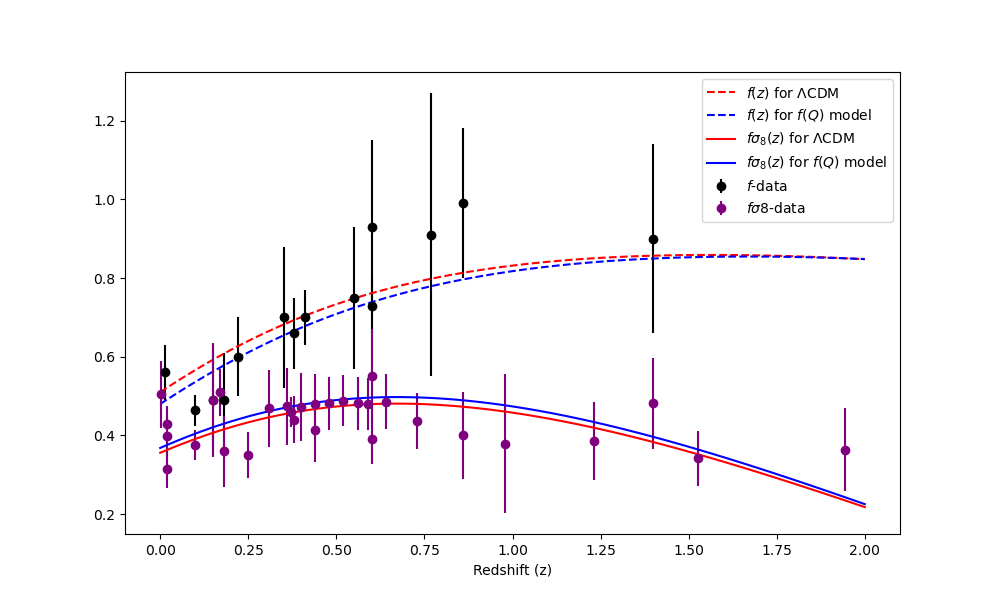}
  \caption{The diagram of the growth rate \(f(z)\) (represented by the red dashed line and blue dashed line) and the redshift-space distortion \(f\sigma_8(z)\) (depicted by the red solid line and blue solid line) as functions of redshift for the $\Lambda$CDM and $f(Q)$-gravity model using Eqs. \eqref{rr1} and \eqref{growth11}.}
    \label{fig:enter-labelx}
\end{figure}
Using the MCMC methods, the corresponding values of the constrained cosmological parameters \( \{\Omega_m, \sigma_8,b_0\}\) are presented in Table. \ref{sphericcase1} using \texttt{f} and \texttt{f}$\sigma_8$ datasets. Using the same datasets, the MCMC simulation results are also provided using $1\sigma$ and $2\sigma$ confidence intervals for both models: $\Lambda$CDM and $f(Q)$ gravity model (\ref{appendix}, Fig. \ref{fig:enter-labelappendex}). By using these parameters' values, we present diagrams of the growth rate \( f(z) \), which indicates how quickly cosmic structures are forming, and the redshift-space distortion \( f\sigma_8(z) \), which combines the growth rate with the clustering amplitude, providing insight into the distribution of matter and the influence of gravity, including contributions from the peculiar velocities of galaxies as well as the Hubble flow, as shown in Fig. \ref{fig:enter-labelx}. {{Although the $\Lambda$CDM model fits the data better, our study shows that the $f(Q)$ gravity model can be a potential alternative, especially in light of existing challenges, such as the issue of $H_0$ and $\sigma_8$ tensions that the $\Lambda$CDM model has not yet been able to resolve.}}

\section{Conclusions}\label{conclusions}
This paper explored the constraints on a toy model of the $f(Q)$ gravitational theory using cosmic measurements, including \texttt{OHD, SNIa, OHD+SNIa}, and large-scale structure (\texttt{f} and \texttt{f}\(\sigma_8\)) datasets. Our analysis of cosmological parameters $\Omega_m$, $\sigma_8$, and $b_0$ indicates that the model \( f(Q) = \alpha\sqrt{Q}+\beta \) is highly effective in reducing the Hubble and \(\sigma_8\) tensions. The lower Hubble parameter value of $69.20^{+4.40}_{\text{--}2.10}$ with the OHD dataset is recorded, while the higher values of $\sigma_8 = 0.827^{+0.030}_{\text{--}0.01}$ are obtained using $f\sigma_8$ datasets. When we considered the SNIa, OHD+SNIa, we found the lower values of $H_0$ for the $f(Q)$ gravity model compared to the $\Lambda$CDM's results, see Table \ref{sphericcase1}. For comparison, the full-mission Planck measurements of the cosmic microwave background (CMB) \cite{aghanim2020planck} yield a Hubble constant $H_0 = (67.4 \pm 0.5) \text{km s}^{-1} \text{Mpc}^{-1}$, a matter density parameter $\Omega_m = 0.315 \pm 0.007$, and amplitude of normalisation of the matter power spectrum $\sigma_8 = 0.811\pm 0.006$. In conclusion, this analysis enhances our understanding of the role of \( f(Q)\) gravity in the late-time accelerated expansion of the Universe and the growth of cosmic structures, helping to evaluate how well this model aligns with or deviates from observational data, thereby providing potential constraints or distinctions from other cosmological theories. In the future, we expect to conduct a comprehensive and in-depth study of cosmological tensions using the latest measurements of CMB data from Planck 2018, gravitational waves, and BAO data from the Dark Energy Survey Instrument together with other sources. 
\section*{References}
\bibliographystyle{iopart-num}
\bibliography{iopart-num}

\appendix
\section{MCMC Results}
\label{appendix}
\begin{figure*}[h!]
    \includegraphics[width=0.35\linewidth]{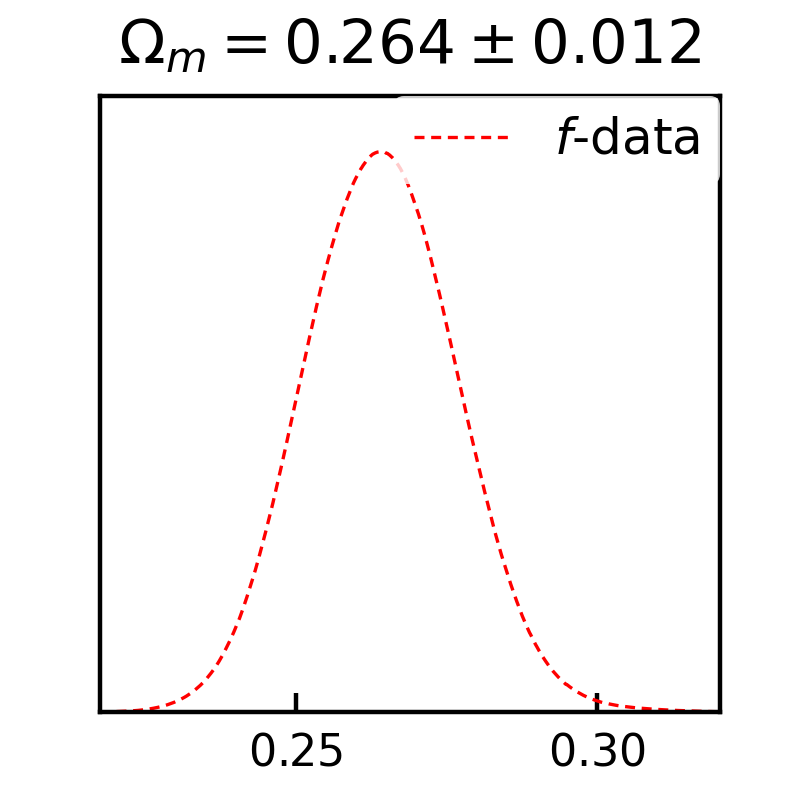}
    \includegraphics[width=0.4\linewidth]{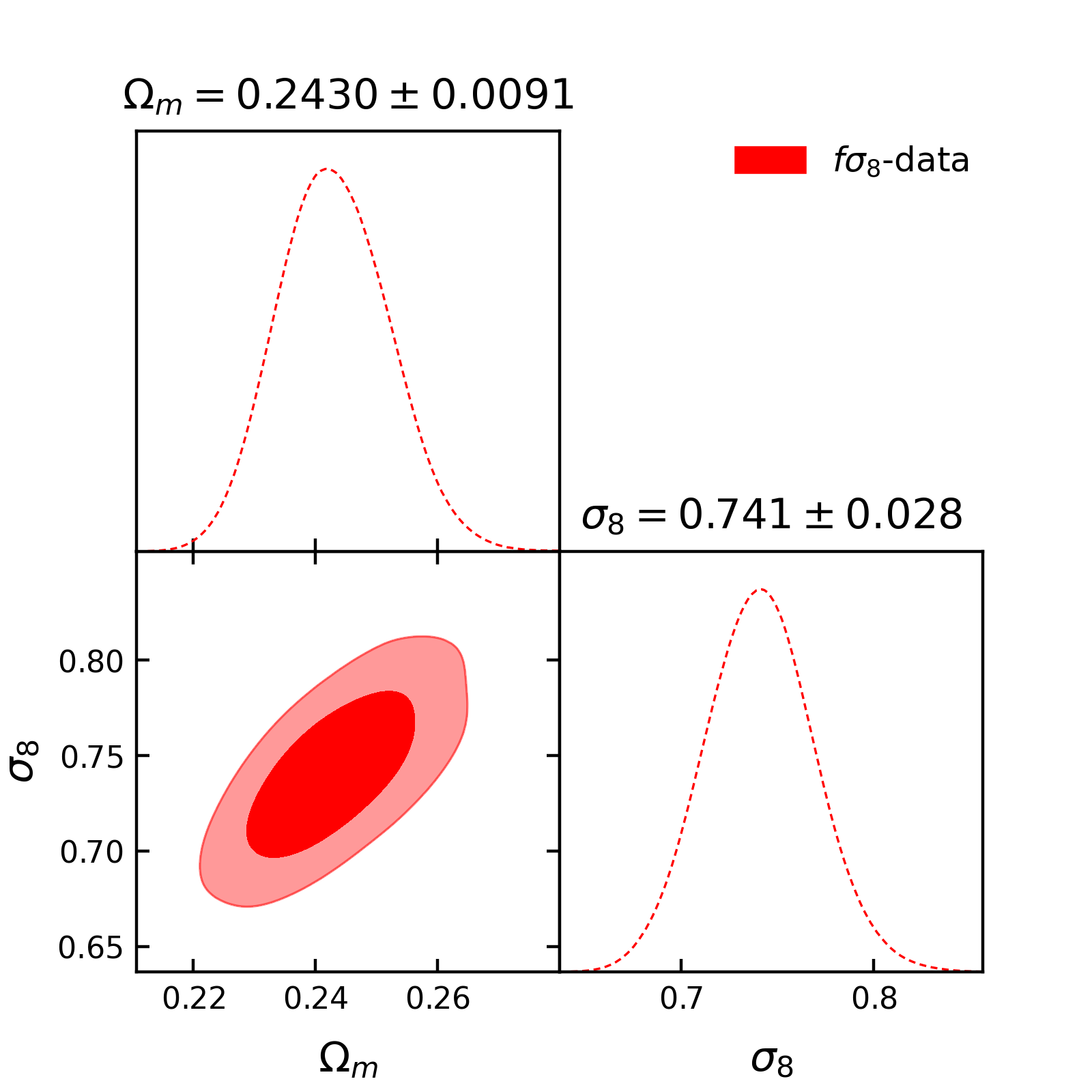}
    \includegraphics[width=0.4\linewidth]{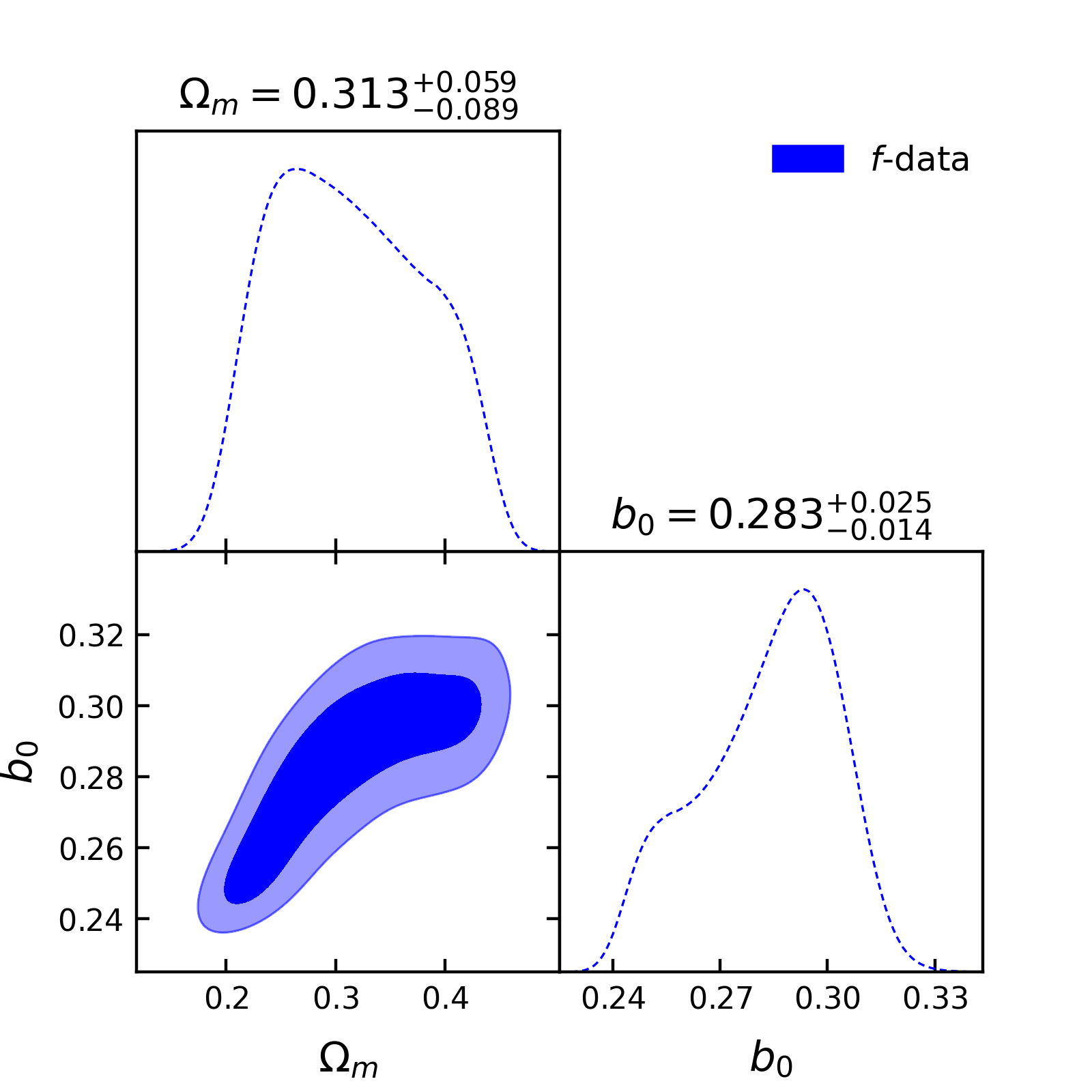}
    \quad~~~~~~~~~~~~~~~~~~~~~~~~~
    \includegraphics[width=0.4\linewidth]{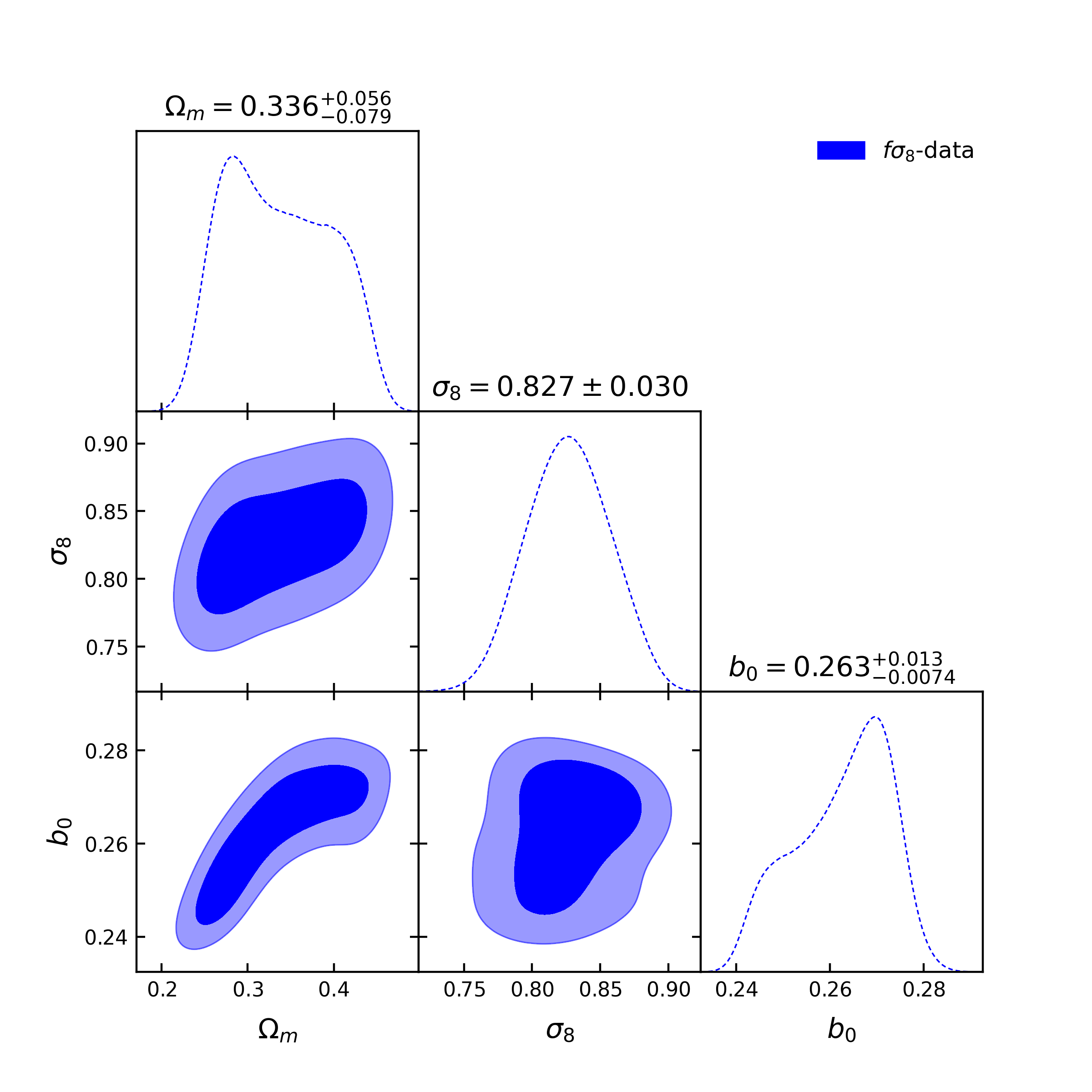}
    \caption{The constrained parameters of $\Omega_m$ and $\sigma_8$ for the $\Lambda$CDM model (top panel) and $\Omega_m$, $\sigma_8$ and $b_0$ for the $f(Q)$ gravity model (bottom panel) using the large-scale structure data  $f$ and $f\sigma_8$. }
    \label{fig:enter-labelappendex}
\end{figure*}
\end{document}

%% file: _06Akalu_revision.bbl
\providecommand{\newblock}{}
\begin{thebibliography}{10}
\expandafter\ifx\csname url\endcsname\relax
  \def\url#1{{\tt #1}}\fi
\expandafter\ifx\csname urlprefix\endcsname\relax\def\urlprefix{URL }\fi
\providecommand{\eprint}[2][]{\url{#2}}

\bibitem{riess1998observational}
Riess A~G, Filippenko A~V, Challis P, Clocchiatti A, Diercks A, Garnavich P~M,
  Gilliland R~L, Hogan C~J, Jha S, Kirshner R~P {\em et~al.\/} 1998 {\em The
  Astronomical Journal\/} {\bf 116} 1009

\bibitem{Perlmutter_1999}
Perlmutter S {\em et~al.\/} 1999 {\em The Astrophysical Journal\/} {\bf 517}
  565 \urlprefix\url{https://dx.doi.org/10.1086/307221}

\bibitem{koivisto2006dark}
Koivisto T and Mota D~F 2006 {\em Physical Review D—Particles, Fields,
  Gravitation, and Cosmology\/} {\bf 73} 083502

\bibitem{daniel2008large}
Daniel S~F {\em et~al.\/} 2008 {\em Physical Review D—Particles, Fields,
  Gravitation, and Cosmology\/} {\bf 77} 103513

\bibitem{eisenstein2005detection}
Eisenstein D~J {\em et~al.\/} 2005 {\em The Astrophysical Journal\/} {\bf 633}
  560

\bibitem{2011MNRAS.417.3101P}
{Percival} W~J {\em et~al.\/} 2011 {\em \mnras\/} {\bf 417} 3101--3102

\bibitem{caldwell2004cosmic}
Caldwell R~R and Doran M 2004 {\em Physical Review D\/} {\bf 69} 103517

\bibitem{huang2006holographic}
Huang Z~Y {\em et~al.\/} 2006 {\em Journal of Cosmology and Astroparticle
  Physics\/} {\bf 2006} 013

\bibitem{harko2011f}
Harko T {\em et~al.\/} 2011 {\em Physical Review D—Particles, Fields,
  Gravitation, and Cosmology\/} {\bf 84} 024020

\bibitem{setare2013can}
Setare M and Mohammadipour N 2013 {\em Journal of Cosmology and Astroparticle
  Physics\/} {\bf 2013} 015

\bibitem{sahlu2020scalar}
Sahlu S, Ntahompagaze J, Abebe A, de~la Cruz-Dombriz {\'A} and Mota D~F 2020
  {\em The European Physical Journal C\/} {\bf 80} 422

\bibitem{sahlua2024cosmology}
Sahlua S, Alfedeelb A~H and Abebe A 2024 {\em arXiv preprint
  arXiv:2406.08303\/}

\bibitem{heisenberg2024review}
Heisenberg L 2024 {\em Physics Reports\/} {\bf 1066} 1--78

\bibitem{mandal2020cosmography}
Mandal S, Wang D and Sahoo P 2020 {\em Physical Review D\/} {\bf 102} 124029

\bibitem{atayde2021can}
Atayde L and Frusciante N 2021 {\em Physical Review D\/} {\bf 104} 064052

\bibitem{jimenez2018coincident}
Jim{\'e}nez J~B, Heisenberg L and Koivisto T 2018 {\em Physical Review D\/}
  {\bf 98} 044048

\bibitem{sokoliuk2023impact}
Sokoliuk O, Arora S, Praharaj S, Baransky A and Sahoo P 2023 {\em Monthly
  Notices of the Royal Astronomical Society\/} {\bf 522} 252--267

\bibitem{frusciante2021signatures}
Frusciante N 2021 {\em Physical Review D\/} {\bf 103} 044021

\bibitem{atayde2023f}
Atayde L and Frusciante N 2023 {\em Physical Review D\/} {\bf 107} 124048

\bibitem{szydlowski2015aic}
Szyd{\l}owski M, Krawiec A, Kurek A and Kamionka M 2015 {\em The European
  Physical Journal C\/} {\bf 75} 5

\bibitem{dunsby1992cosmological}
Dunsby P, Bruni M and Ellis G 1992 {\em Astrophys. J\/} {\bf 395} 34

\bibitem{sahlua2024structure}
Sahlu S, de~la Cruz-Dombriz {\'A} and Abebe A 2024 {\em arXiv preprint
  arXiv:2405.07361\/}

\bibitem{abebe2013large}
Abebe A, de~la Cruz-Dombriz A and Dunsby P~K 2013 {\em Physical Review D\/}
  {\bf 88} 044050

\bibitem{springel2006large}
Springel V, Frenk C~S and White S~D 2006 {\em nature\/} {\bf 440} 1137--1144

\bibitem{yu2018hubble}
Yu H, Ratra B and Wang F~Y 2018 {\em The Astrophysical Journal\/} {\bf 856} 3

\bibitem{dixit2023observational}
Dixit A {\em et~al.\/} 2023 {\em Indian Journal of Physics\/} {\bf 97}
  3695--3705

\bibitem{scolnic2018complete}
Scolnic D~M {\em et~al.\/} 2018 {\em The Astrophysical Journal\/} {\bf 859} 101

\bibitem{avila2022inferring}
Avila F, Bernui A, Bonilla A and Nunes R~C 2022 {\em The European Physical
  Journal C\/} {\bf 82} 594

\bibitem{hough2020viability}
Hough R, Abebe A and Ferreira S 2020 {\em The European Physical Journal C\/}
  {\bf 80} 787

\bibitem{lewis2019getdist}
Lewis A 2019 {\em arXiv preprint arXiv:1910.13970\/}

\bibitem{staicova2022constraining}
Staicova D and Benisty D 2022 {\em Astronomy \& Astrophysics\/} {\bf 668} A135

\bibitem{aghanim2020planck}
Aghanim N, Akrami Y, Ashdown M, Aumont J, Baccigalupi C, Ballardini M, Banday
  A~J, Barreiro R, Bartolo N, Basak S {\em et~al.\/} 2020 {\em Astronomy \&
  Astrophysics\/} {\bf 641} A6

\end{thebibliography}
